\documentclass[sigconf]{acmart}

\acmConference{}{}{}

\renewcommand\footnotetextcopyrightpermission[1]{}
\usepackage{stfloats}

\title{Whispering Water: Materializing Human-AI Dialogue as Interactive Ripples}

\author{Ruipeng Wang}
\authornote{These authors contributed equally to this work.}
\affiliation{%
  \institution{MIT Media Lab, Critical Matter Group}
  \city{Cambridge}
  \state{MA}
  \country{USA}
}
\email{ruipengw@mit.edu}
\author{Tawab Safi}
\authornotemark[1]
\affiliation{%
  \institution{MIT Media Lab, Viral Communications Group}
  \city{Cambridge}
  \state{MA}
  \country{USA}
}
\email{safi7092@mit.edu}
\author{Yunge Wen}
\authornotemark[1]
\affiliation{%
  \institution{MIT Media Lab, New York University}
  \city{Cambridge}
  \state{MA}
  \country{USA}
}
\email{yungew@mit.edu}
\author{Christina Cunningham}
\affiliation{%
  \institution{Massachusetts Institute of Technology}
  \city{Cambridge}
  \state{MA}
  \country{USA}
}
\email{Chrimc@mit.edu}
\author{Hoi Ling Tang}
\affiliation{%
  \institution{Harvard University}
  \city{Cambridge}
  \state{MA}
  \country{USA}
}
\email{helentang@gsd.harvard.edu}
\author{Behnaz Farahi}
\affiliation{%
  \institution{MIT Media Lab, Critical Matter Group}
  \city{Cambridge}
  \state{MA}
  \country{USA}
}
\email{behnaz\_f@media.mit.edu}

\begin{abstract}
Water has long served as a recipient of human confession across cultures. We present \textit{Whispering Water}, an interactive installation that materializes human-AI dialogue through cymatic patterns on water. Participants confess to a water surface, triggering a four-phase ritual: confession, contemplation, response, and release. Speech sentiment is translated into excitation frequencies that prime the water's physical state, while semantic content enters a multi-agent system of heterogeneous LLMs whose identities emerge through situated discourse. A novel algorithm decomposes synthesized speech into harmonic components via logarithmic spacing and Bark-scale mapping, reconstructing machine voices as physical wave superpositions. The installation explores emotional self-exploration through sensory-rich, ritually framed human-AI interaction.
\end{abstract}

\begin{document}

\begin{teaserfigure}
  \centering
  \includegraphics[width=\textwidth]{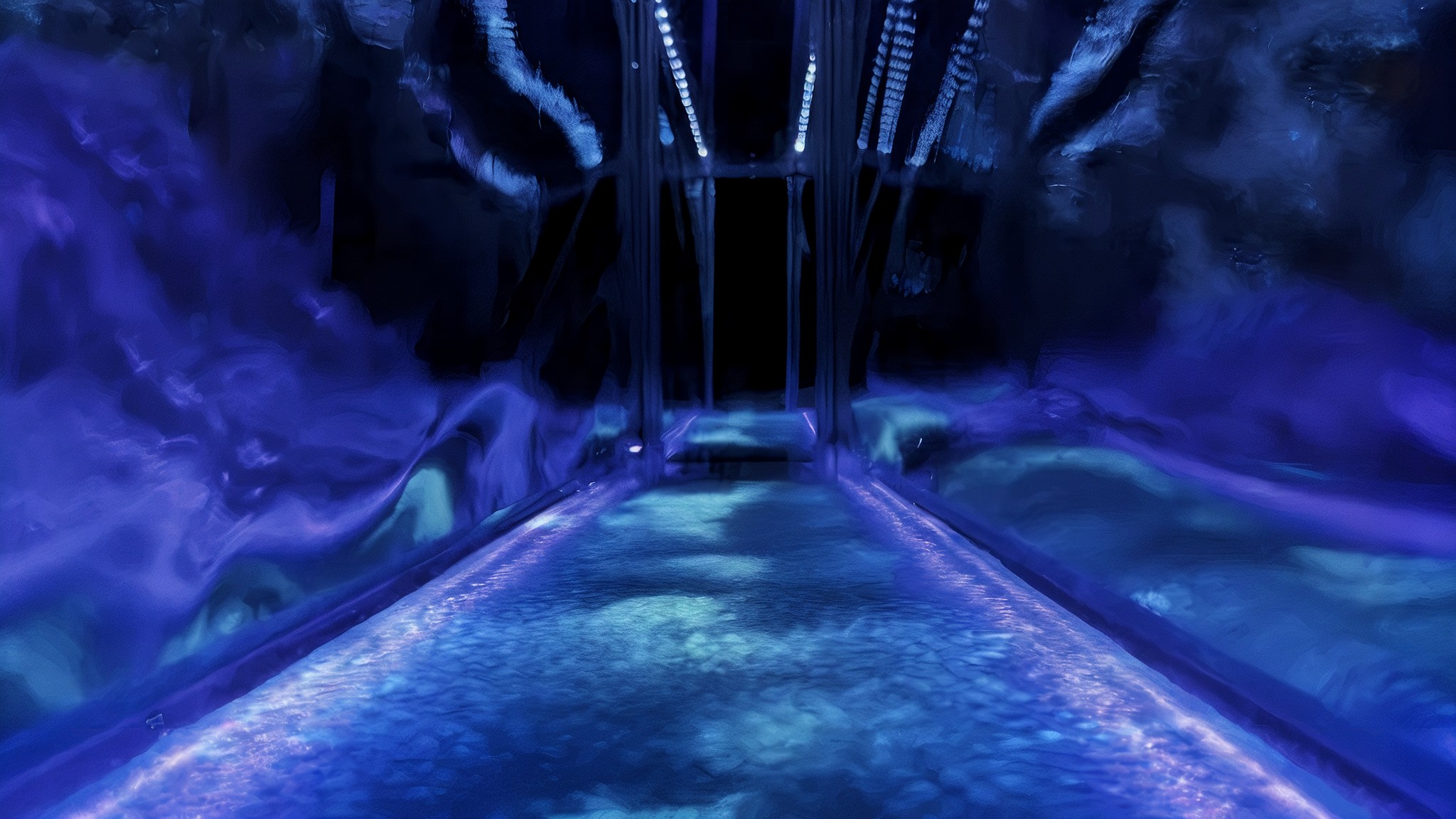}
  \caption{We introduce \textit{Whispering Water}, an interactive installation that materializes human-AI dialogue through cymatic patterns on water. A participant's confession to the water surface triggers a multi-agent dialogue among six heterogeneous LLMs. Each agent's synthesized speech is decomposed into harmonic components and reconstructed as mechanical vibrations through six subwoofers, generating the cymatic interference patterns visible on the water surface. By rendering machine reasoning as emergent physical phenomena, the installation reframes human-AI interaction as a sensory and ritual experience, opening new possibilities for emotional self-exploration through material interfaces.}
  \label{fig:teaser}
\end{teaserfigure}

\maketitle

\begin{figure*}

    \centering
    \includegraphics[width=1\linewidth]{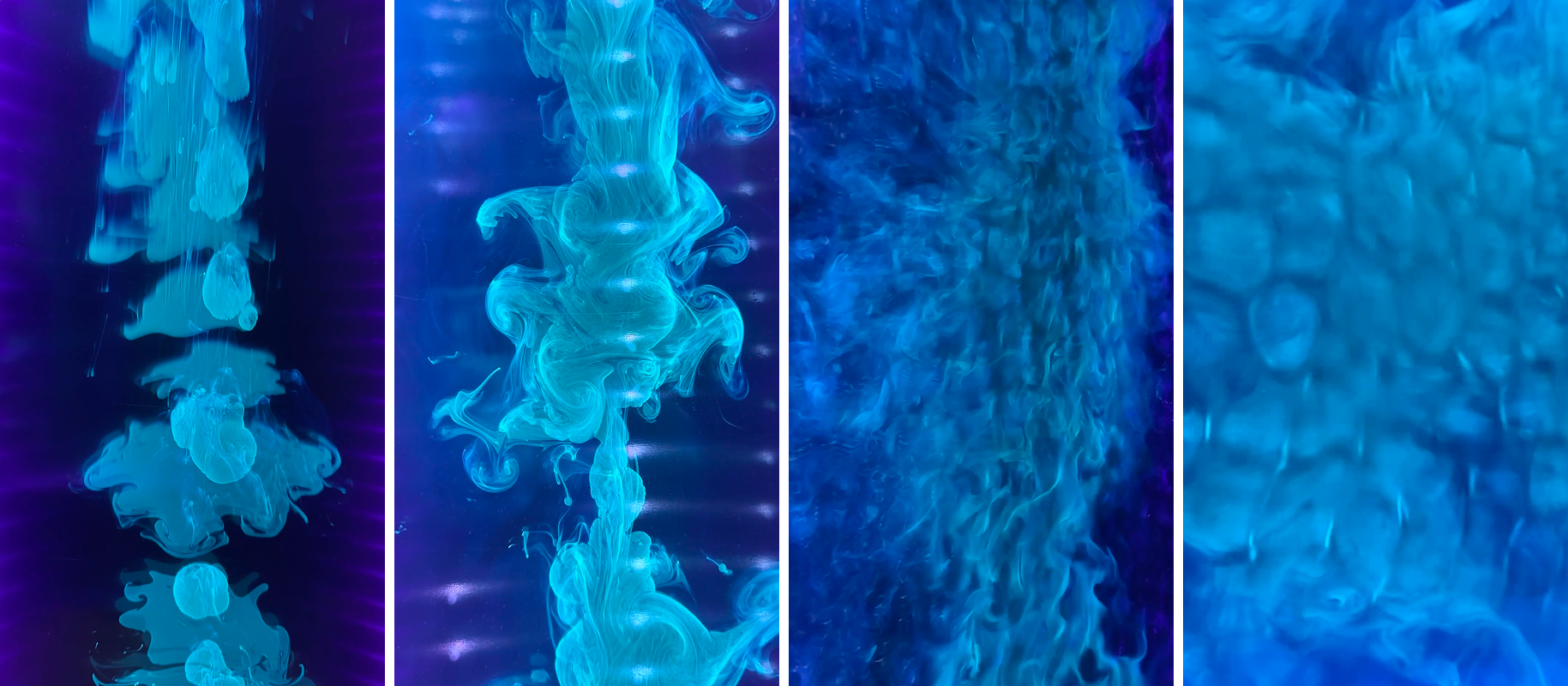}

    \caption{\textbf{Cymatic patterns across ritual stages.} UV-reactive dye and UV LEDs render surface formations visible. (1) Confession: the surface remains still. (2) Contemplation: standing waves form at frequencies corresponding to detected emotional sentiment. (3) Response: agents speak sequentially then simultaneously, their wave components interfering to produce complex overlapping patterns. (4) Release: the summarizer agent's concluding statement is decomposed into six simultaneous waves, filling the tank as a collective final response.}
    \label{fig:ripple_comparison}
\end{figure*}

\section{\textbf{Introduction}}

People have long whispered to water. Celtic peoples believed spirits in sacred wells could grant wishes spoken aloud~\cite{celtic}. In Japanese misogi, practitioners chant while immersed, asking kami to cleanse impurities~\cite{Japan}. Greek hydromancy interpreted ripples as divine messages~\cite{Greek}. Daoist philosophy understands water as yin, receptive yet powerful through softness~\cite{Tao}. Across these traditions, water functions as a listener: a medium that receives vulnerability, responds through subtle material change, and returns to stillness without judgment. The confession spoken to water is not merely symbolic. It is a phenomenological act in which the human voice becomes physically inscribed in a yielding surface, visible for a moment before dissolving.

Contemporary human-AI interaction inherits none of this. Conversational interfaces are optimized for efficiency, coherence, and task resolution. They render machine reasoning as text or synthesized voice, formats that are legible but affectively thin. The intimacy of confession, the weight of unburdening, the ceremony of release: these dimensions of human emotional life find no foothold in the interaction paradigms dominant today. We ask what it would mean to design an AI interface that takes confession seriously as its interactional frame, and that responds not through language alone but through the physics of a material the human body already understands as receptive.

We present \textit{Whispering Water}, an interactive installation that materializes human-AI dialogue through cymatic patterns on water. Participants speak to a water surface; their voice undergoes sentiment analysis to prime the water's physical state, while semantic content is forwarded to six AI agents engaged in parallel dialogue. The agents respond to both the participant and one another, producing moments of convergence and divergence before reconciliation. Each agent's synthesized speech is decomposed into harmonic components and reconstructed through acoustic superposition, translating machine voices into cymatic patterns: from voice to vibration, sentiment to frequency, linguistic meaning to material form. The interaction unfolds across four phases of confession, contemplation, response, and release, structuring the exchange as a ritual rather than a query.

\begin{figure*}
    \centering
    \includegraphics[width=1\linewidth]{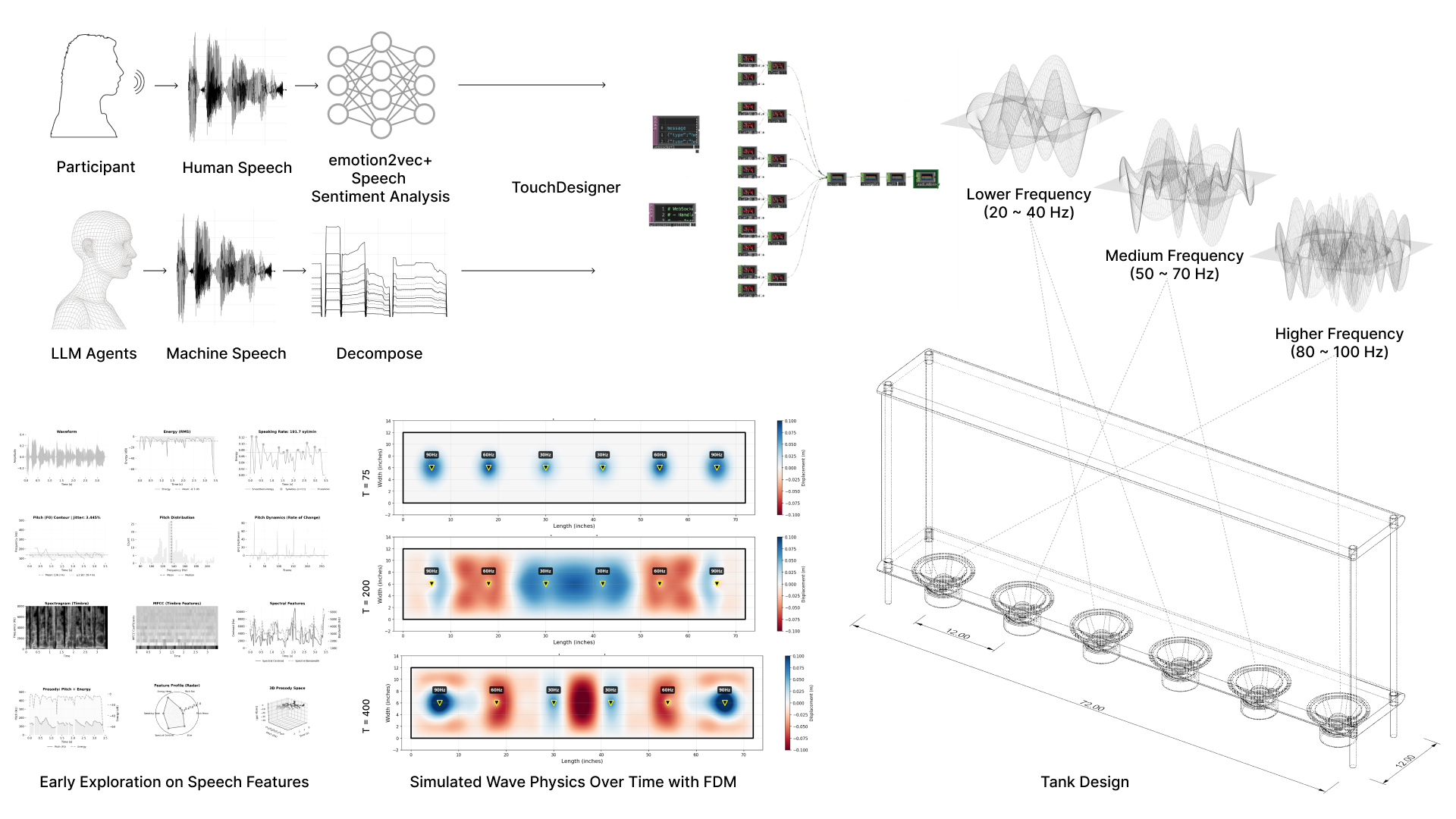}

    \caption{\textbf{System overview and installation design.} Human speech and LLM agent outputs are processed through two parallel pipelines routed through TouchDesigner to six subwoofers. The upper pipeline maps emotional sentiment via emotion2vec+ to excitation frequencies; the lower decomposes machine speech into harmonic components using the Bark-scale algorithm (Section~\ref{sec:tech}). Subwoofers are arranged in three bands: central (20--40~Hz), intermediate (50--70~Hz), and outer (80--100~Hz). FDM simulations ($T = 75, 200, 480$) show how superposed vibrations produce evolving interference patterns, informing the tank design.}
    
    \label{fig:installation_design}
\end{figure*}

\section{\textbf{Related Works}}

\textbf{Water and Cymatics as Artistic Medium.
}Water has been extensively explored as a responsive material in interactive art, across forms including droplets~\cite{droplet1, droplet2, droplet3}, bubbles~\cite{bubble1, bubble2}, streams~\cite{stream}, and mist~\cite{mist}. Cymatic approaches have made water's acoustic responsiveness visible: Turczan's \textit{Mirror Speak} drives water surfaces with low-frequency vibrations~\cite{mirror_speak}; \textit{Sonic Water} invited participants to generate surface patterns through voice~\cite{sonic_water}; Bai et al.'s \textit{Sonic Shower} explored sonification of water as a speculative multisensory interface~\cite{bai2025sonicshower}. While these works establish water as an expressive acoustic medium, they do not connect surface phenomena to machine reasoning or dialogue.

\textbf{Materializing AI Output.}
Recent work has explored giving AI systems physical or perceptual presence. Naseck et al. manifested a generative AI music system through a kinetic sculpture, translating musical decisions into expressive movement to bridge the gap between machine output and audience perception~\cite{naseck2025physical}. Similarly, CRIA embodied AI emotional states through a digital human avatar, using voice synthesis and facial animation to make machine affect tangible~\cite{cai2025cria}. These works share our interest in rendering opaque AI processes as perceptible phenomena, though they operate in visual and sonic registers rather than material ones.

\textbf{Emotional Interfaces and Ritual Interaction.}
Affective computing has long sought to align interface behavior with human emotional states~\cite{picard1997affective}. More recent work examines interfaces that support emotional self-exploration rather than task completion, designing for ambiguity, intimacy, and reflection. Bakhtin's dialogism~\cite{bakhtin1981dialogic} and Suchman's situated action theory~\cite{suchman1987plans} inform our treatment of agent identity as emergent through discourse. Our work extends these threads by grounding emotional interaction in physical ritual: confession as a culturally resonant act that frames the human-AI exchange as meaningful rather than transactional.

\begin{figure*}
    \centering
    \includegraphics[width=1\linewidth]{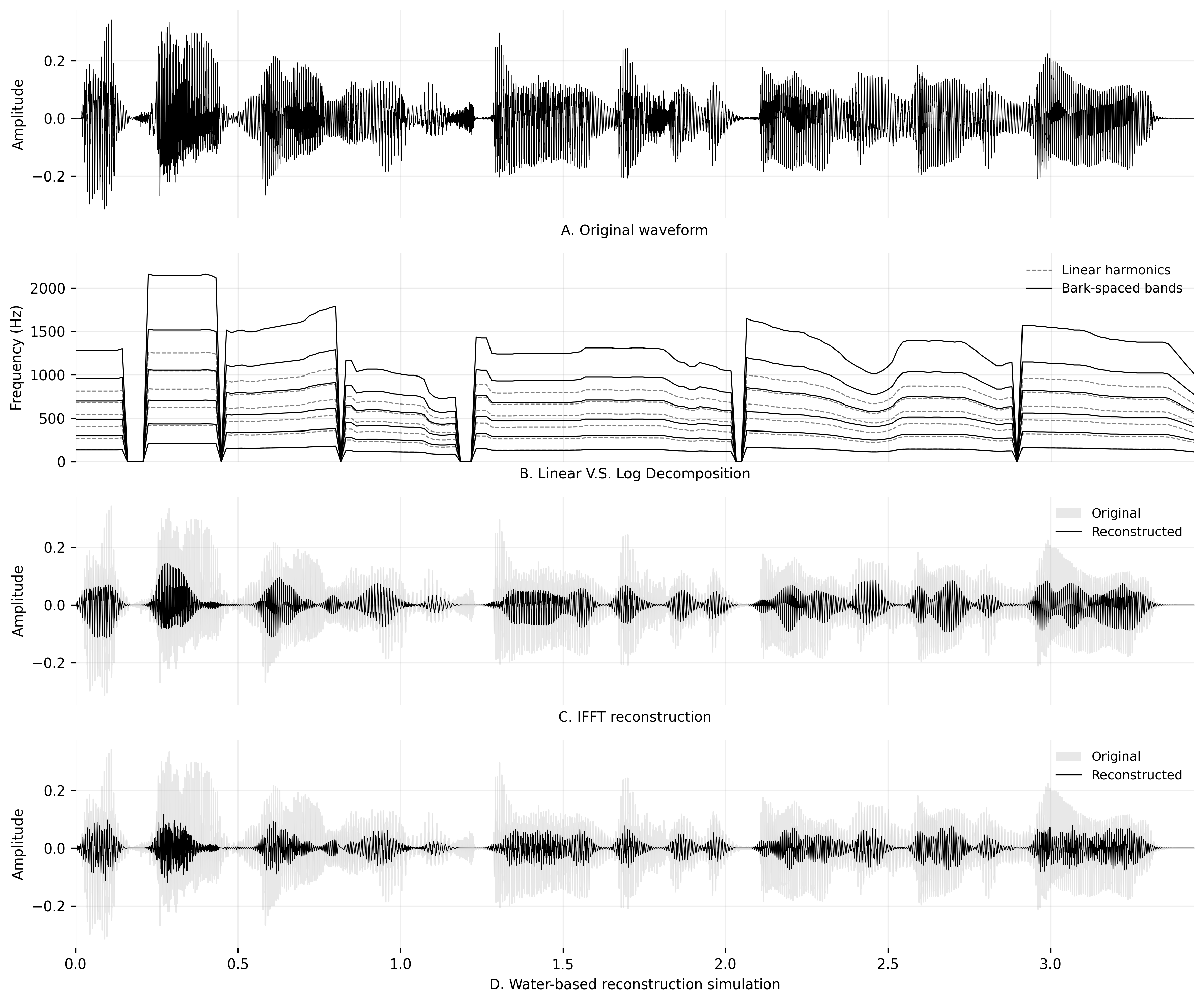}
    \caption{\textbf{Machine speech decomposition and reconstruction. }(A) Original waveform of approximately 3-second machine-synthesized audio. (B) Comparison between linear and log decomposition methods, showing linear harmonics (grey dashed lines) versus Bark-spaced frequency bands (black lines). (C) IFFT reconstruction using linear harmonic components. (D) Water-based reconstruction simulation using log-spaced (Bark scale) method. The visualization demonstrates that the log + Bark approach covers a broader frequency range aligned with human auditory perception, resulting in richer information content in the reconstructed signal.}
    \label{fig:speech_translation}
\end{figure*}

\section{\textbf{System Design}}

Whispering Water distills the cultural phenomenon of confession to water into a four-phase ritual: confession, contemplation, response, and release. This sequence shifts the participant's role from speaker to witness as water becomes the primary site of meaning.

\textbf{Confession.} Participants lean forward to speak close to the water surface, adopting a posture of intimacy associated with prayer and confession. The system records the voice for up to 15 seconds while the water remains still.

\textbf{Contemplation.} Overhead lighting slowly rises as a standing wave forms on the surface at a frequency reflecting the extracted emotional sentiment, priming the water before any linguistic response begins.

\textbf{Response.} Six AI agents engage across four rounds of dialogue: independent reflection, peer selection, targeted response, and collective synthesis. Each agent's voice is decomposed into wave components and reconstructed in water sequentially; waves interfere with residual patterns from prior speakers, accumulating a material record of the exchange. In the final round, a summarizer agent's concluding statement is decomposed into six simultaneous waves that fill the entire tank.

\textbf{Release.} Vibration diminishes until the surface returns to stillness. All input and responses are encrypted for privacy, and the confession dissolves with the water.

The installation is an elongated metal frame (72 × 12 × 32 inches) enclosed with Mylar film and mirrored Plexiglass. The tunnel geometry creates an intimate space for disclosure; mirrored surfaces fold the viewer into the water while the Mylar membrane couples vibration directly into the vessel. UV-reactive dye and UV LEDs make surface patterns visible, rendering the inaudible seen. Six subwoofers mechanically coupled to the frame propagate low-frequency vibrations into the water across three frequency bands; their placement follows standing wave physics, which concentrates cymatic activity at the basin's midpoint, directly informing the spatial logic of the algorithm described in Section~\ref{sec:tech}.

\section{\textbf{Technical Implementation}}
\label{sec:tech}

\subsection{\textbf{Sentiment Analysis of Human Speech}}

Unlike human speech, synthesized machine voices tend toward uniform rhythm and constrained emotional range~\cite{tts}. We therefore use the participant's vocal quality as a primer, establishing an initial cymatic state during the contemplation phase before agent interaction begins. We deploy emotion2vec+ to analyze speech acoustics, capturing pitch, energy, speaking rate, timbre, and prosody. Emotional output is translated into excitation frequencies: low (20--40~Hz) for slow undulations associated with calm or melancholic affect; mid-range (50--70~Hz) for regular standing waves; and high (80--100~Hz) for dense, energetic surface activity corresponding to aroused or anxious states. These frequencies are emitted simultaneously across all six subwoofers, establishing the water's material disposition before the agents respond.

\subsection{\textbf{Multi-agent Conversation}}

Agent-to-agent communication in multi-agent systems typically assigns role identities through task-based allocation~\cite{agent_capability} or frameworks like the BDI model~\cite{agent_bdi}. However, Bakhtin's dialogism~\cite{bakhtin1981dialogic} and Suchman's situated action theory~\cite{suchman1987plans} suggest that meaning and agency emerge through interaction rather than residing in individuals. Our system treats agent roles as dynamically constituted through discourse rather than predefined.

The agent pool comprises multiple untuned LLMs (Claude Sonnet 4.5, GPT-4.5, Gemini 2.0, etc.) to introduce heterogeneity in reasoning patterns and linguistic styles. The participant's speech is transcribed via ASR and forwarded to six agents processing in parallel. Each agent produces a brief response (150 characters) while selecting a voice persona encoded in JSON. All agents execute concurrently, with outputs interleaved through asynchronous queue multiplexing.

Rather than assigning fixed vocal identities, agents dynamically select TTS personas based on content and tone, allowing vocal identity to emerge from discourse. Personas draw from expressive voice profiles in ElevenLabs spanning diverse ages, genders, and registers. In early rounds, agents speak sequentially to maintain perceptual clarity; in the final round, they speak simultaneously, creating interference patterns that physically manifest the convergence and divergence of their interpretations.

\subsection{\textbf{Speech-to-Wave Translation}}

The core technical contribution of this work is an algorithm that decomposes synthesized speech into frequency components and reconstructs them physically through water. Each agent's voice is mapped to one to six wave components played through the six subwoofers; when multiple agents speak simultaneously, their waves superpose in the water vessel.

\textbf{Decomposition.} Speech signals are transformed via the Short-Time Fourier Transform (STFT), yielding 257 frequency bins (0--8000~Hz). Frequencies between 85--255~Hz correspond to the fundamental frequency ($f_0$); higher harmonics encode timbre and intelligibility. Rather than selecting linear multiples of $f_0$, we apply logarithmic spacing from $f_0$ to $8 \times f_0$ (Equation~\ref{eq:log}) and map components to the Bark scale, a psychoacoustic measure based on auditory critical bands (Equation~\ref{eq:bark})~\cite{bark}, deriving six wave components per speech signal.

\begin{equation}
f_i = f_0 \times 8^{i/5}, \quad i = 0, 1, 2, 3, 4, 5
\label{eq:log}
\end{equation}

\begin{equation}
\text{Bark}(f_i) = 13 \cdot \arctan(0.00076 \cdot f_i) + 3.5 \cdot \arctan\left(\left(\frac{f_i}{7500}\right)^2\right)
\label{eq:bark}
\end{equation}

\textbf{Reconstruction.} With phase preserved, STFT-decomposed audio can be reconstructed through IFFT with minimal loss, a process that occurs physically as sound waves superpose in air. Water behaves differently: its speed of sound (${\sim}1500$~m/s) produces longer wavelengths ($\lambda = v/f$), and low frequencies (20--100~Hz) propagate stably while exhibiting cymatic behavior. We assign components to three frequency bands across the six subwoofers: central subwoofers (3 and 4) at 20--40~Hz, intermediate (2 and 5) at 50--70~Hz, and outer (1 and 6) at 80--100~Hz. Each wave component is dynamically adjusted to its assigned band, ensuring spatially distributed interference patterns that vary with each agent's vocal content.

\section{\textbf{Discussion}}

Whispering Water proposes water as an epistemically active medium that returns human expression through physical law rather than representational logic. Where most AI interfaces render machine reasoning as text or voice, this installation renders it as interference: visible, transient, and irreproducible. Situating agent identity in discourse rather than role assignment reflects a commitment to emergence over control; the resulting dialogue is not always harmonious, and we argue this tension is generative. The ritual structure frames the participant not as a user optimizing for an outcome but as a confessor entering a ceremony whose resolution is material rather than linguistic.

The speech-to-wave translation is bounded by the cymatic range of water (20--100~Hz), and future iterations might explore hybrid transducer configurations or embrace frequency loss as part of the installation's meaning. More broadly, this work speculates toward interfaces in which AI reasoning is not displayed but physically enacted, asking what it means for a machine to respond not through language but through the world.

\bibliographystyle{ACM-Reference-Format}
\bibliography{short}
\end{document}